\documentclass[journal,draftcls,onecolumn,12pt,twoside]{IEEEtran}
\usepackage{algorithm} 
\usepackage{algpseudocode}
\usepackage{graphicx}
\usepackage{amssymb}
\usepackage{amsfonts}
\usepackage[cmex10]{amsmath}
\usepackage{dsfont}
\usepackage{float}
\usepackage{url}
\usepackage{bm}
\usepackage{hhline}
\usepackage{enumerate}
\usepackage{pgf,tikz}
\usepackage{mathrsfs}
\usepackage[framed,numbered,autolinebreaks,useliterate]{mcode}
\usetikzlibrary{arrows}
\usepackage{array}
\usepackage{tabulary}
\newcolumntype{K}[1]{>{\centering\arraybackslash}p{#1}}

 \usepackage[colorlinks=true]{hyperref}
\hypersetup{urlcolor=blue, citecolor=blue,linkcolor=purple}

\IEEEoverridecommandlockouts
\newtheorem{Theorem}{Theorem}
\newtheorem{Proposition}{Proposition}
\newtheorem{Corollary}{Corollary
}
\newtheorem{Definition}{Definition}

\newtheorem{theorem}{Theorem}

\newtheorem{lemma}[theorem]{Lemma}

\makeatletter
\let\@@pmod\pmod
\DeclareRobustCommand{\pmod}{\@ifstar\@pmods\@@pmod}
\def\@pmods#1{\mkern4mu({\operator@font mod}\mkern 6mu#1)}
\makeatother

\usepackage{tikz}
\usetikzlibrary{shapes,arrows}
\usepackage{verbatim}
\usetikzlibrary{arrows}

\tikzstyle{startstop} = [rectangle, rounded corners, minimum width=1.5cm, minimum height=0.65cm,text centered, draw=black, fill=blue!3]
\tikzstyle{io} = [trapezium, trapezium left angle=70, trapezium right angle=110, minimum width=3cm, minimum height=1cm, text centered, draw=black, fill=blue!30]
\tikzstyle{process} = [rectangle, minimum width=3cm, minimum height=1cm, text centered, draw=black, fill=orange!30]
\tikzstyle{decision} = [diamond, minimum width=3cm, minimum height=1cm, text centered, draw=black, fill=green!30]
\tikzstyle{arrow} = [thick,->,>=stealth]

\ifCLASSINFOpdf
\else
\fi
\begin{document}
\title{Practical Encoder and Decoder for Power Constrained  QC-LDPC lattices}
\author{Hassan Khodaiemehr, Mohammad-Reza~Sadeghi and Amin~Sakzad
\thanks{H. Khodaiemehr  and M.~R Sadeghi are with the Department
of Mathematics and Computer Science, Amirkabir University of Technology (Tehran Polytechnic), Tehran, Iran (emails: h.khodaiemehr@aut.ac.ir and msadeghi@aut.ac.ir).

A. Sakzad is  with Clayton School of IT,  Monash University, Victoria, Australia
(e-mail: amin.sakzad@monash.edu).

Part of this work has been presented in \cite{IWCIT, IWCIT2015}.
}}


\maketitle
\begin{abstract}
LDPC lattices were the first family of lattices that equipped with iterative decoding algorithms under which they perform very well in high dimensions.
In this paper, we introduce quasi cyclic low density parity check
(QC-LDPC) lattices as a special case of LDPC lattices with one binary QC-LDPC code as their underlying code. These lattices are obtained from Construction A of lattices providing us to encode them efficiently using shift registers.
To benefit from an encoder with linear complexity in dimension of the lattice, we obtain the generator matrix of these lattices in \emph{quasi cyclic} form.
We provide a low-complexity decoding algorithm of QC-LDPC lattices based on sum product algorithm.
To design lattice codes,  QC-LDPC lattices are combined with  nested lattice shaping
that uses the Voronoi region of a sublattice for code shaping. The shaping
gain and shaping loss of our lattice codes with dimensions $40$, $50$ and $60$ using
an optimal quantizer, are presented.
Consequently, we establish a family of lattice codes that perform practically close to the sphere bound.
\end{abstract}
\section{Introduction}
\PARstart{P}{oltyrev}~\cite{polytrev} suggests and investigates coding without restriction for infinite arrays such as lattices on the additive white Gaussian noise (AWGN) channel. That is a communication without power constraints. In such a communication system, instead of the coding rate and capacity, two new concepts are defined: normalized logarithmic density (NLD) and generalized capacity $C_{\infty}$. Forney  et al. \cite{forneyspherebound} proved theoretically, the existence of sphere-bound-achieving and capacity-achieving lattices via Construction D. They also established the concept of volume-to-noise (VNR) ratio as a parameter for measuring the efficiency of lattices. Therefore, generalized capacity for lattices means the existence of  a lattice with high enough dimension $n$ that enables transmission with arbitrary small error probability whenever VNR approaches~$1$. In addition, it can be shown~\cite{polytrev} that this error probability  is bounded away from zero when $\mbox{VNR}<1$. A capacity-achieving lattice can raise to a capacity-achieving lattice code by selecting a proper shaping region~\cite{erez,urbanke}.

The search for sphere-bound-achieving and capacity-achieving lattices and lattice codes has begun with~\cite{sadeghi}. Low density parity check (LDPC) lattices are those that have sparse parity check matrices. These lattices were introduced first by Sadeghi {\em et al.}~\cite{sadeghi}. In this class of lattices, a set of nested binary LDPC codes along with Construction D' are used to generate lattices with sparse parity check matrices. Another class of lattices, so-called low density lattice codes (LDLC) introduced and investigated in~\cite{3} and \cite{LDLC}. Turbo lattices employed Construction D along with turbo codes to achieve capacity gains~\cite{sakzad10}. Integer low-density lattices based on construction A which are known as LDA lattices \cite{19} and polar lattices~\cite{polar}, are another families of lattices with practical decoding methods. By applying non-binary LDPC codes and Construction A, LDA lattice of dimension $10000$ have obtained an error performance within $0.7$dB of Poltyrev's limit \cite{19,LDAnew}. These lattices are shown to be capacity-achieving even without using dithering technique \cite{LDA2016}.

All the above mentioned high-dimensional lattices share many common properties. For example, they all (except turbo lattices) exploit the parity check matrix of the lattice and employ a relevant message-passing decoding algorithm. However, they are different in some aspects especially when it comes to practical implementations.
In fact, owning a simple and low complexity encoding method is an advantage in the implementation considerations for a family of lattices. In order to design a simple low complexity lattice encoder with low-storage requirement, the first prerequisite is obtaining a generator matrix of the lattice in a special form like circular or quasi-cyclic structure with small integer components. Hence,  the main contributions of this work are as follows.

\begin{itemize}\itemsep -2pt
\item We introduce and investigate quasi cyclic (QC) LDPC lattices which are special case of LDPC lattices \cite{sadeghi}. This family of lattices are supported with both a practical decoder as well as an efficient lattice encoder. Different encoding approaches based on parallel, serial and two-stage shift-register-adder-accumulator (SRAA) circuits can be adapted. Furthermore, the computational complexity of these algorithms with respect to clock-cycles and flip-flops are determined.
\item We establish QC-LDPC lattice codes to be used in power constraint Gaussian channels. In order to obtain finite constellations from these lattices, we employ Voronoi shaping method. We compute the shaping gain of these lattice codes at low dimensions. Finally, we conduct simulations providing numerical results to reveal the effectiveness of QC-LDPC lattices in terms of fundamental coding gain and symbol error probability.
\end{itemize}


This paper is organized as follows. In Section~\ref{Preliminaries}, we provide some preliminaries about lattices. The definition of  QC-LDPC lattices is also presented in this section.  In Section~\ref{Encoding}, the generator matrix of  QC-LDPC lattices in different cases is obtained.
In Section~\ref{decod}, we propose a practical method for encoding of QC-LDPC lattices.  We also present two decoding methods for  QC-LDPC lattices based on sum-product decoding algorithm of LDPC codes. In Section~\ref{shaping_sec}, Voronoi  shaping method is applied to  QC-LDPC lattices and shaping gain/loss of these lattices at low dimensions are computed approximately. In Section~\ref{Simulations}, we  present the simulation results of the error decoding performance. Section~\ref{ConclusionandSuggestionforLaterResearch} contains the concluding remarks.

\textbf{Notation}: Matrices and vectors are denoted by bold upper
and lower case letters. The $i^{th}$ element of a vector $\mathbf{a}$ is denoted
by $a_i$ and the $(i,j)^{th}$ entry of a matrix $\mathbf{A}$ is denoted by
$A_{i,j}$ unless otherwise stated. $[\,\,]^t$ denotes the transposition for vectors and matrices.
\section{Preliminaries}\label{Preliminaries}
\subsection{Lattices}\label{A.Lattices}
A discrete, additive subgroup $\Lambda$ of the $m$-dimensional real space $\mathbb{R}^m$ is called a \emph{lattice}.
Every lattice $\Lambda$ has a basis $\mathcal{B}=\{{\mathbf b}_1,\ldots,{\mathbf b}_n\}\subseteq\mathbb{R}^m$, where every ${\mathbf x}\in\Lambda$ can be represented as an integer linear combination of the vectors in $\mathcal{B}$.
The rank of the lattice is $n$ and its dimension is $m$. If $n = m$, the lattice is called a full-rank lattice. In this paper, we consider full-rank lattices. The matrix $\mathbf{M}$ with $\mathbf{b}_1,\ldots, \mathbf{b}_n$ as rows, is a generator matrix
for the lattice. The matrix $\mathbf{G} = \mathbf{M}\mathbf{M}^t$ is  the Gram matrix for
the lattice. The determinant of the lattice, denoted by $\det(\Lambda)$, is  the determinant of the matrix $\mathbf{G}$ and
the volume of the lattice is defined as $\textrm{vol}(\Lambda) =\sqrt{\det(\mathbf{G})}$.
 A Voronoi cell $\mathcal{V}(\mathbf{x})$ is the set of those points of $\mathbb{R}^n$ that are at least as close to $\mathbf{x}$ as to any other point in $\Lambda$. We call the Voronoi region associated with the origin,
the fundamental Voronoi region of $\Lambda$, denoted by $\mathcal{V}$ or $\mathcal{V}(\Lambda)$.

The \emph{normalized volume} of an $n$-dimensional lattice $\Lambda$ is defined as $\textrm{vol}(\Lambda)^{\frac{2}{n}}$~\cite{forneyspherebound}. This volume may be regarded as the volume of $\Lambda$ per two dimensions. Suppose that the points of a lattice $\Lambda$ are sent over an unconstrained additive white Gaussian noise (AWGN)~\cite{polytrev} channel, with noise variance $\sigma^2$. Let the vector $\mathbf{x}\in\Lambda$ be transmitted over the unconstrained AWGN channel, then the received vector $\mathbf{r}$ can be written as $\mathbf{r}=\mathbf{x}+\mathbf{e}$,
where $\mathbf{e}=(e_1,\ldots,e_n)$ is the error term and its components are
independently and identically distributed  (i.i.d.) with $\mathcal{N}(0,\sigma^2)$.
The \emph{volume-to-noise ratio} (VNR) of lattice $\Lambda$ is
\begin{equation}\label{VNR}
{\mbox {VNR}}=\frac{\textrm{vol}(\Lambda)^{\frac{2}{n}}}{2\pi e\sigma^2}.
\end{equation}
For a large $n$, the VNR is the ratio of the normalized volume of $\Lambda$ to the normalized volume of a noise sphere of squared radius $n\sigma^2$ which is defined as generalized signal-to-noise ratio (SNR) in~\cite{sadeghi} and $\alpha^2$ in~\cite{forneyspherebound}.
The probability of  correct decoding is
\begin{equation}~\label{pe}
P_c(\Lambda)=\frac{1}{(\sigma\sqrt{2\pi})^n}\int_{\mathcal{V}(\mathbf{x})}e^{\frac{-\|\mathbf{t}\|^2}{2\sigma^2}}d\mathbf{t},
\end{equation}
where $\|{\mathbf x}\|$ is the Euclidean norm of $\mathbf{x}$.

A \emph{lattice constellation} $C(\Lambda,\mathcal{R})=(\Lambda +\mathbf{t})\cap \mathcal{R}$ is a finite set of points in a lattice translate $\Lambda + \mathbf{t}$ that lies within a compact bounding region $\mathcal{R}$ of the $n$-dimensional real space $\mathbb{R}^n$. The key geometric properties
of the region $\mathcal{R}$ are its \emph{volume} $\textrm{vol}(\mathcal{R})$ and the \emph{average energy} $P(\mathcal{R})$ per dimension of a uniform probability density function over $\mathcal{R}$ (see, e.g., \cite{1} and \cite{2}):
\begin{equation}\label{eq1}
    P(\mathcal{R})=\int_\mathcal{R} \frac{(\|\mathbf{x} \|^2/n)d\mathbf{x}}{\textrm{vol}(\mathcal{R})}.
\end{equation}
The \emph{normalized second moment} of $\mathcal{R}$ is
\begin{equation}\label{eq2}
    G(\mathcal{R})=\frac{P(\mathcal{R})}{\textrm{vol}(\mathcal{R})^{2/n}}.
\end{equation}
The normalized second moment of any $n$-cube centered at the origin is $1/12$. The \emph{shaping gain} $\gamma_s(\mathcal{R})$ of $\mathcal{R}$, measures the decrease in average energy of $\mathcal{R}$ relative to a baseline region, namely, an interval $[-d_0/2, d_0/2]$ or an $n$-cube $[-d_0/2, d_0/2]^n$, where $d_0$ is related to the $\textrm{vol}(\mathcal{R})$ \cite{2}. The shaping gain of $\mathcal{R}$ is
\begin{equation}\label{eq3}
    \gamma_s(\mathcal{R})=\frac{\textrm{vol}(\mathcal{R})^{2/n}}{12 P(\mathcal{R})}=\frac{1}{12G(\mathcal{R})}.
\end{equation}
The optimum $n$-dimensional shaping region is an $n$-sphere~\cite{1}. The key geometrical parameters of an $n$-sphere $(=\otimes)$
of radius $r$ for an even $n$ are \cite{2}:
\begin{eqnarray}
  \textrm{vol}(\otimes) &=& \frac{(\pi r^2)^{n/2}}{(n/2)!}, \label{sphereV}\\
  P(\otimes) &=& \frac{r^2}{n+2}\label{sphereP},\\
  G(\otimes) &=& \frac{P(\otimes)}{\textrm{vol}(\otimes)^{2/n}}=\frac{((n/2)!)^{2/n}}{\pi(n+2)}. \label{sphereG}
\end{eqnarray}
The shaping gain of an $n$-sphere is a function of the dimension $n$. For example its value for dimension $100$ is about $1.37$dB.  When $n$ approaches infinity  we see that the shaping gain approaches the \emph{ultimate shaping gain} $\pi e /6$ ($1.53$dB). The \emph{shaping loss} $\lambda_{s} (\mathcal{R})$ of a shaping region $\mathcal{R}$ with respect to an $n$-dimensional sphere, where $n$ is even, based on (\ref{sphereV})-(\ref{sphereG}),  is~\cite{3}:
\begin{equation}\label{eq6}
    \lambda_{s}(\mathcal{R})=\frac{G(\mathcal{R})}{G(\otimes)}=\frac{\pi (n+2)G(\mathcal{R})}{\Gamma(\frac{n}{2}+1)^{2/n}}.
\end{equation}
The shaping loss is greater than or equal to $1$.
\subsection{LDPC lattices}\label{LDPC Lattices}
There exist many ways to construct lattices based on codes \cite{2}.
Assume that $\mathcal{C}$ is a linear code over $\mathbb{F}_p$ where $p$ is a prime number, i.e. $\mathcal{C}\subseteq\mathbb{F}_{p}^n$. A lattice $\Lambda$ based on Construction A~\cite{2} can be derived from $\mathcal{C}$ as follows
\begin{equation}\label{constA}
\Lambda=p\mathbb{Z}^n+\epsilon\left(\mathcal{C}\right),
\end{equation}
where $\epsilon\colon\mathbb{F}_{p}^n\rightarrow\mathbb{R}^n$ is the embedding function. In this work, we are particularly interested in lattices with $p=2$.

Construction D' converts a set of parity checks defined by a family of nested codes $\mathcal{C}_0\supseteq \mathcal{C}_1\supseteq \cdots \supseteq \mathcal{C}_a$, into congruences for a lattice~\cite{2}. The number $a+1$ is called the level of the construction.
An LDPC lattice $\Lambda\subset \mathbb{Z}^n$ can be constructed from Construction D' and a number of nested binary LDPC codes.
More detail about the structure and decoding of these lattices can be found in \cite{sadeghi}. If we consider one code as underlying code of Construction D', which means $a=0$, Construction A is obtained~\cite[Proposition 1]{IWCIT}. In this case, Construction A LDPC lattices or 1-level LDPC lattices \cite{IWCIT} are obtained. In this paper, we refer to them as LDPC lattices without mentioning the level of the construction.
Now, we introduce a new subclass of LDPC lattices for which we present efficient encoding and decoding procedures in the sequel.
\begin{Definition}
A  QC-LDPC lattice $\Lambda$ is a lattice based on Construction A along with one binary  QC-LDPC code $\mathcal{C}$ as its underlying code. Equivalently, $\mathbf{x}\in\mathbb{Z}^n$ is in $\Lambda$ if and only if $\mathbf{H}_{qc}\mathbf{x}^t=\mathbf{0} \pmod{2}$, where $\mathbf{H}_{qc}$ is a quasi cyclic parity check matrix of $\mathcal{C}$.
\end{Definition}

In the rest of this paper, $\mathbf{H}_{qc}$ denotes the parity check matrix of a QC-LDPC lattice or equivalently the parity check matrix of its underlying code. Note that when $\mathbf{H}_{qc}$ is the parity check matrix of a lattices, the operations are performed over $\mathbb{R}$, while if $\mathbf{H}_{qc}$ denotes the parity check of a code, the operations are done over a binary field $\mathbb{F}_2$. The dimension of the QC-LDPC lattices is denoted by $tb$ instead of $n$, following the notations of~\cite{17}.
%
%
\section{Generator matrix of QC-LDPC lattices}\label{Encoding}
In this section, we address the problem of finding the generator matrix of QC-LDPC lattices. The generator matrix is needed not only for encoding, also for the computation of lattice shaping gain. First, we address the general case, i.e., when the considered  lattice is an arbitrary LDPC lattice. Then, we obtain the quasi cyclic generator matrix of QC-LDPC lattices which is divided  into two different cases.
%
As mentioned before, the considered LDPC lattices in this paper can be  represented as a Construction A lattices.
The generator matrix of Construction A lattice $\Lambda$ using the underlying code $\mathcal{C}$ is of the form \cite{2}:
\begin{eqnarray}\label{eq12}
  \mathbf{G}_{\Lambda} &=& \left[
                    \begin{array}{cc}
                      \mathbf{I}_{k}& \mathbf{P}_{k\times (n-k)} \\
                      \mathbf{0}_{(n-k)\times k} & 2\mathbf{I}_{n-k} \\
                    \end{array}
                  \right],
\end{eqnarray}
\noindent
where $\mathbf{G}_{\mathcal{C}}=\left[
                           \begin{array}{cc}
                              \mathbf{I}_k & \mathbf{P}\\
                           \end{array}
                         \right]$
is the generator matrix of $\mathcal{C}$ in the systematic form, $k$ is the rank of $\mathcal{C}$ and $n$ is the code length of $\mathcal{C}$.
The matrices $\mathbf{I}_k$ and $\mathbf{0}_k$, are identity and the all zero square matrices of size $k$, respectively.
The generator matrix of QC-LDPC lattice $\Lambda$ with underlying code $\mathcal{C}$, can be obtained by permuting  the columns of $\mathbf{G}_{\Lambda}$ in (\ref{eq12}).
\begin{Proposition}\label{prop2}
Let $\mathbf{H}_{qc}$ be the parity check matrix of a QC-LDPC code $\mathcal{C}$ with code length $n=tb$ and dimension $k=cb$, where $c,t$ and $b$ are positive integers. Let us consider $\mathcal{C}$ as the underlying code of the QC-LDPC lattice $\Lambda$. Then, the generator matrix of $\Lambda$ will be of the form $\mathbf{G}_{\Lambda}\mathbf{T}$, where $\mathbf{G}_{\Lambda}$ is given in (\ref{eq12}) and $\mathbf{T}$ is a permutation matrix that permutes the columns of $\mathbf{H}_{qc}$  so that the last $n-k$ columns of the obtained matrix be independent. Moreover, $\mathbf{H}_{qc}$ is the parity check matrix of  $\Lambda$.
\begin{IEEEproof}
The proof is trivial.
\end{IEEEproof}
\end{Proposition}
It should be noted that, when the parity check matrix $\mathbf{H}'$ of the underlying code $\mathcal{C}$ is not in quasi cyclic form, we can consider $\mathbf{H}=\mathbf{H}'\mathbf{T}$ and $\mathbf{G}_{\Lambda}$ in (\ref{eq12}) as the parity check matrix and the generator matrix of a Construction A lattice $\Lambda$, respectively. In this case, we disregard  the matrix $\mathbf{T}$ in Proposition~\ref{prop2} and $\mathbf{H}$ serves as the parity check matrix of $\Lambda$. When $\mathbf{H}'$ is in quasi cyclic form, like Proposition~\ref{prop2}, we can not do this, because $\mathbf{H}'\mathbf{T}$ is no longer quasi cyclic and using $\mathbf{H}'\mathbf{T}$ for decoding increases the complexity. Moreover,  using the proposed generator matrix in Proposition~\ref{prop2} for QC-LDPC lattices entails high storage requirements, which increases the encoding computational complexity to $O(n^2)$. For large $n$, this incurs high computational encoding costs, which is considered as one of the main practical implementation challenges. In the sequel, we present the generator matrix of QC-LDPC lattices in quasi cyclic form no matter if the parity check matrix $\mathbf{H}_{qc}$ of the underlying QC-LDPC code contains a  full-rank quasi cyclic sub-matrix or it is rank-deficient. At one hand, if $\mathbf{H}_{qc}$ itself is full-rank, the encoding complexity is related to the number of polynomials used to generate $\mathbf{H}_{qc}$, which is much less compared to $n^2$. On the other hand, if $\mathbf{H}_{qc}$ is rank-deficient, we again represent the generator matrix of the obtained QC-LDPC lattice in a format which includes only circulant matrices. This again significantly reduce the encoding complexity of such lattices.
\subsection{QC-LDPC lattices with an invertible QC sub-matrix in $\mathbf{H}_{qc}$}
For the sake of implementing the encoding operation with low complexity, we consider  QC-LDPC lattices. The authors of \cite{17} and \cite{18} proposed an efficient encoder for QC-LDPC codes. Their proposed encoding is simplified by obtaining the generator matrix of QC-LDPC codes  in partial quasi-cyclic form, comprising an identity matrix, a parity generator matrix, a zero matrix and a remainder matrix. Let $\mathbf{H}_{qc}$  of size $cb\times tb$, with $c\leq t$, be the parity check matrix of the underlying code $\mathcal{C}$. Let $\mathbf{H}_{qc}$ has full rank, $r=cb$, and there exists a  $cb\times cb$ quasi cyclic sub-matrix $\mathbf{D}$ in $\mathbf{H}_{qc}$  with rank $r$, i.e., $\mathbf{D}$ is an invertible quasi cyclic matrix over $\mathbb{F}_2$.  Then, we obtain the quasi cyclic generator matrix of $\mathcal{C}$ in the following systematic form \cite{17}
\begin{equation}\label{gen}
  \mathbf{G}_{qc}=\left[
           \begin{array}{cccccccc}
             \mathbf{I} & \mathbf{0} & \cdots & \mathbf{0} & | & \mathbf{G}_{1,1}  & \cdots & \mathbf{G}_{1,c} \\
             \mathbf{0} & \mathbf{I} & \cdots & \mathbf{0} & | & \mathbf{G}_{2,1}  & \cdots & \mathbf{G}_{2,c} \\
             \vdots & \vdots & \ddots & \vdots             & | & \vdots & \ddots & \vdots \\
             \mathbf{0} & \mathbf{0} & \cdots & \mathbf{I} & | & \mathbf{G}_{t-c,1}  & \cdots & \mathbf{G}_{t-c,c} \\
           \end{array}
         \right],
\end{equation}
where each $\mathbf{G}_{i,j}$, with $1\leq i\leq t-c$ and $1\leq j\leq c$, is a $b\times b$ circulant matrix. In this case, the generator matrix of the QC-LDPC lattice $\Lambda$ that is obtained from QC-LDPC code $\mathcal{C}$ with generator matrix $\mathbf{G}_{qc}$ in~(\ref{gen}), is of the form~(\ref{eq12}) by replacing $\mathbf{G}_{\mathcal{C}}=\left[
                           \begin{array}{cc}
                              \mathbf{I}_k & \mathbf{P}\\
                           \end{array}
                         \right]$ with $\mathbf{G}_{qc}$.
\subsection{QC-LDPC lattices with rank-deficient $\mathbf{H}_{qc}$}
In most cases, the quasi cyclic matrix $\mathbf{H}_{qc}$ is rank-deficient and we can not obtain a quasi cyclic sub-matrix $ \mathbf{D}$ inside $\mathbf{H}_{qc}$. It should be noted that  we can not use the elementary row operations to eliminate the dependent rows of $\mathbf{H}_{qc}$ and get a full-rank sub-matrix of $\mathbf{H}_{qc}$, because we want to exploit from the quasi cyclic structure of the parity check matrix to simplify the decoding and encoding operations. Indeed, using the elementary row operations give us a full-rank sub-matrix of $\mathbf{H}_{qc}$ that is not quasi cyclic and it is useless, which does not fit to our framework.

Let $\mathbf{e}_i$ be a row vector with a single $1$ in the $i^{th}$ position and $0$ elsewhere.
When $\mathbf{H}_{qc}$ is rank deficient,  the generator matrix of QC-LDPC lattices can be obtained as follows. Let $r$ and $cb$, with $r<cb$, be the rank and the number of the rows of $\mathbf{H}_{qc}$, respectively. Find  the  positions of  $r$ independent columns of $\mathbf{H}_{qc}$ and consider them as $\left\{i_1,\ldots ,i_r\right\}\subset \left\{1,\ldots,n\right\}$. Next, it is proved that by stacking the rows of the generator matrix of underlying code $\mathcal{C}$ and the vectors $2\mathbf{e}_{i_j}$, for $1\leq j\leq r$, into a matrix,  we obtain the generator matrix of QC-LDPC lattices. In the previous case, these positions were the last $n-k$ positions.

\noindent In the sequel,
assume that  $r<cb$ or $r=cb$ but there does not exist a full-rank quasi cyclic sub-matrix like $\mathbf{D}$ in $\mathbf{H}_{qc}$ with rank $r$.  In this case, we first find the least number of columns of circulants in $\mathbf{H}_{qc}$, say $l$, with $c \leq l\leq t$, such that these $l$ columns of $b\times b$ circulants form a $c\times l$ subarray $\mathbf{D}^{*}$, whose rank is equal to the rank $r$ of  $\mathbf{H}_{qc}$. Indeed, $\mathbf{D}^{*}$ is a $cb\times lb$ quasi cyclic submatrix of $\mathbf{H}_{qc}$ with rank $r$. We permute the columns of circulants of $\mathbf{H}_{qc}$ to
form a new $c\times t$ array $\mathbf{H}^{*}_{qc}$ of $b\times b$ circulants, such that the last $l$ columns of circulants form the array $\mathbf{D}^{*}$. Then, the  generator matrix of code $\mathcal{C}$ with this parity check matrix, is a $(tb-r)\times tb$ matrix, and has the following form \cite{17}:
\begin{equation}\label{gen2}
  \mathbf{G}_{qc}^{*}=\left[
               \begin{array}{c|c}
                 \mathbf{G}^t & \mathbf{Q}^t
               \end{array}
             \right]^t,
\end{equation}
which consists of two sub-matrices $\mathbf{G}$ and $\mathbf{Q}$.
The  sub-matrix $\mathbf{G}$ is a $(t-l)\times t$ array of $b\times b$ blocks of the form (\ref{gen}).
The sub-matrix $\mathbf{Q}$ of $\mathbf{G}_{qc}^{*}$ is an $(lb-r)\times tb$ matrix whose
rows are linearly independent, and also linearly independent of the rows of the sub-matrix $\mathbf{G}$ of $\mathbf{G}_{qc}^{*}$. The matrix $\mathbf{Q}$ has the following form:
\begin{equation}\label{Qmat}
 \mathbf{Q}= \left[
    \begin{array}{ccccccc}
      \mathbf{0}_{1,1} & \cdots & \mathbf{0}_{1,t-l} & | & \mathbf{Q}_{1,1} & \cdots & \mathbf{Q}_{1,l} \\
      \vdots & \ddots & \vdots & | & \vdots & \ddots & \vdots \\
      \mathbf{0}_{l,1} & \cdots & \mathbf{0}_{l,t-l} & | & \mathbf{Q}_{l,1} & \cdots & \mathbf{Q}_{l,l} \\
    \end{array}
  \right],
\end{equation}
where each $\mathbf{0}_{i,k}$ is a $d_i\times b$ zero matrix for $1\leq i\leq b$ and $1\leq k\leq t-l$, $d_i$'s will be introduced next, and $\mathbf{Q}_{i,j}$ is a  matrix over $\mathbb{F}_2$ for $1\leq i,j\leq l$. Each nonzero sub-matrix $\mathbf{Q}_{i,j}$ is a partial circulant matrix obtained by cyclically shifting the first row of $\mathbf{Q}_{i,j} $ one place to the right $d_i-1$ times. Therefore, $\mathbf{Q}$ also has a partial circulant\footnote{A partial circulant matrix has the following form $$\mathbf{Q}=\left[
               \begin{array}{cccc}
                 \mathbf{Q}_{1,1} & \mathbf{Q}_{1,2} & \cdots & \mathbf{Q}_{1,t} \\
                 \vdots & \vdots & \ddots & \vdots \\
                 \mathbf{Q}_{c,1} & \mathbf{Q}_{c,2} & \cdots & \mathbf{Q}_{c,t} \\ \cline{1-4}
                 \mathbf{Q}_{c+1,1} & \mathbf{Q}_{c+1,2} & \cdots & \mathbf{Q}_{c+1,t} \\
                 \vdots & \vdots & \ddots & \vdots \\
                 \mathbf{Q}_{c+l,1} & \mathbf{Q}_{c+l,2} & \cdots & \mathbf{Q}_{c+l,t} \\
               \end{array}
             \right],
$$ where $\mathbf{Q}_{i,j}$, for $1\leq i\leq c$ and $1\leq j\leq t$, is a $b\times b$ circulant matrix, for $c\leq i\leq c+l$ and $1\leq j\leq t$, $\mathbf{Q}_{i,j}$ is a $d_i\times b$ matrix with $d_i<b$, where each row vector of it is rotated one element to the right relative to its preceding row vector.} structure.

Considering the columns $ib+1,\ldots, (i+1)b$ of $\mathbf{D}^{*}$, for $0\leq i\leq l-1$, the columns $ib,\ldots, ib+d_i$ form a linearly dependent set. Thus, the sequence $d_1,d_2,\ldots , d_l$ shows the number of linearly dependent columns in the $1^{st}, 2^{nd},\cdots , l^{th}$ columns of circulants in $\mathbf{D}^{*}$, respectively. Therefore, $\sum_{i=1}^l d_i=lb-r$.
For $1\leq i \leq l$,
let $\mathbf{q}_i=(\mathbf{0},\ldots , \mathbf{0},q_{i,1},\ldots , q_{i,lb})$ be the first row of the submatrix
$\left[
   \begin{array}{cccccc}
     \mathbf{0}_{i,1} & \cdots & \mathbf{0}_{i,t-l} & \mathbf{Q}_{i,1} & \cdots & \mathbf{Q}_{i,l} \\
   \end{array}
 \right]
$, which is the $i^{th}$ row of $\mathbf{Q}$ and its first $(t-l)b$ components are zero.  The $lb-r$ bits of the vector $\mathbf{w}_i\triangleq(q_{i,1},\ldots , q_{i,lb})$, corresponding to the positions of linearly dependent columns of $\mathbf{D}^{*}$, are known as follows $(\mathbf{0}_{d_1},\ldots,\mathbf{0}_{d_{i-1}},\mathbf{u}_i,\mathbf{0}_{d_{i+1}},\ldots,\mathbf{0}_{d_l})$,
where  $\mathbf{0}_{d_s}$ is the all zero row vector of size $d_s$, for $1\leq s \leq l$, and $\mathbf{u}_i$ is a unit $d_i$-tuple, i.e., a row vector of length $d_i$ with $1$ in its first position and zero in the other positions. Based on the structure of $\mathbf{w}_i$, the number of unknown components of $\mathbf{w}_i$ are $r$, the same as the rank of  $\mathbf{D}^{*}$. For each $1\leq j\leq l$ define $\bar{d_j}=b-d_j$. Note that $\sum_{i=0}^l \bar{d_i}=lb-(lb-r)=r$. By solving $\mathbf{D}^{*}\cdot \mathbf{w}_i^t=\mathbf{0}$ over $\mathbb{F}_2$,  we find $\mathbf{w}_i$ and accordingly $\mathbf{q}_i$, for $1\leq i\leq l$. Thus, $\mathbf{Q}_{i,j}$, for $1\leq i,j \leq l$, are obtained by  cyclic shift of the following vector $d_i-1$ times to the right
\begin{equation}\label{q_{ij}}
  \mathbf{q}_{i,j}=\left(\delta_{i,j},\overbrace{0,\ldots,0}^{d_j-1},q_{jb+d_j+1,1},\ldots ,q_{(j+1)b,1}\right),
\end{equation}
where $\delta_{i,j}=1$ if $i=j$ and $\delta_{i,j}=0$ otherwise. Consider first $d_i$ columns of $\mathbf{Q}_{i,i}$ for $1\leq i\leq l$, which is a $d_i\times d_i$ lower triangular matrix with $1$ on the components of the main diagonal. The columns of this matrix are linearly independent over $\mathbb{F}_2$.

\begin{Theorem}\label{prop3}
Let $\mathcal{C}$ has a generator matrix as given in (\ref{gen2}). Then, the generator matrix of the  QC-LDPC lattice $\Lambda=\mathcal{C}+2\mathbb{Z}^n$ is of the form
\begin{eqnarray}\label{lattice_gen_main}
\mathbf{G}_{\Lambda}=\left[
\begin{array}{c|c}
\mathbf{G}_{qc}^{*t} &
\mathbf{R}^t
\end{array}
\right]^t,
\end{eqnarray}
where sub-matrix $\mathbf{R}$ is defined in (\ref{latticegen}).
\begin{equation}
\label{latticegen}
\mathbf{R}= \left[
    \begin{array}{c|c}
      \mathbf{0}_{r\times (t-l)b} &  \begin{array}{ccccccccc}
\mathbf{0}_{\bar{d_1}\times d_1} & 2I_{\bar{d_1}\times \bar{d_1}} & \mathbf{0}_{\bar{d_1}\times d_2} & \mathbf{0}_{\bar{d_1}\times \bar{d_2}} &      \mathbf{0}_{\bar{d_1}\times d_3}& \mathbf{0}_{\bar{d_1}\times \bar{d_3}}&\cdots & \mathbf{0}_{\bar{d_1}\times d_l} & \mathbf{0}_{\bar{d_1}\times \bar{d_l}} \\
\mathbf{0}_{\bar{d_2}\times d_1}  & \mathbf{0}_{\bar{d_2}\times \bar{d_1}} &  \mathbf{0}_{\bar{d_2}\times d_2} & 2I_{\bar{d_2}\times \bar{d_2}}  &      \mathbf{0}_{\bar{d_2}\times d_3}& \mathbf{0}_{\bar{d_2}\times \bar{d_3}}&\cdots & \mathbf{0}_{\bar{d_2}\times d_l} & \mathbf{0}_{\bar{d_2}\times \bar{d_l}} \\
\mathbf{0}_{\bar{d_3}\times d_1} & \mathbf{0}_{\bar{d_3}\times \bar{d_1}} & \mathbf{0}_{\bar{d_3}\times d_2} & \mathbf{0}_{\bar{d_3}\times \bar{d_2}}  &  \mathbf{0}_{\bar{d_3}\times d_3} &2 I_{\bar{d_3}\times \bar{d_3}}&\cdots & \mathbf{0}_{\bar{d_3}\times d_l} & \mathbf{0}_{\bar{d_3}\times \bar{d_l}}  \\
\vdots &\vdots & \vdots & \vdots  & \vdots &\vdots &\ddots & \vdots &\vdots \\
\mathbf{0}_{\bar{d_l}\times d_1} &\mathbf{0}_{\bar{d_l}\times \bar{d_1}} & \mathbf{0}_{\bar{d_l}\times d_2} & \mathbf{0}_{\bar{d_l}\times \bar{d_2}}  &\mathbf{0}_{\bar{d_l}\times d_3} &\mathbf{0}_{\bar{d_l}\times \bar{d_3}}&\cdots &\mathbf{0}_{\bar{d_l}\times d_l} &2I_{\bar{d_l}\times \bar{d_l}} \\
    \end{array}        \\
    \end{array}
  \right].
\end{equation}
\begin{IEEEproof}
We must find a set of independent vectors over $\mathbb{Z}$ in $\Lambda$ and show that they  generate every vector in $\Lambda$. We put these vectors as the rows of the generator matrix of $\Lambda$. The proof is complicated and it is based  on induction. It needs some definitions and lemmas which are all presented in Appendices~\ref{app0} and~\ref{appB}.
\end{IEEEproof}
\end{Theorem}

We conclude that the generator matrix of the QC-LDPC lattices can be obtained in each of the aforementioned two  cases, i.e., when the generator matrix of underlying code can be expressed in the quasi cyclic systematic form, which is a rare case, or the case that it can only be expressed  in the partial quasi-cyclic form. Based on  the proof of Theorem~\ref{prop3}, we have the following corollary.
\begin{Corollary}\label{cor2}
If $\Lambda$ is a  QC-LDPC lattice with parity check matrix $\mathbf{H}_{qc}^{*}$ and generator matrix of the form~(\ref{lattice_gen_main}), then
\begin{equation}\label{det_QC}
\det(\Lambda)=2^r,
\end{equation}
where $r=\textrm{rank}(\mathbf{H}_{qc}^{*})$.
\end{Corollary}
\section{Encoding and decoding of QC-LDPC lattices}\label{decod}

In this paper, our concentration is to find lattices with good error correcting capabilities  and low encoding-decoding complexity. The symbol error rate (SER) of the uncoded layer $p\mathbb{Z}$ of the Construction A lattices  at $\textrm{VNR}=0$dB has the following form \cite{polytrev,19}
\begin{equation}\label{err}
  P_e(p\mathbb{Z})= 2Q\left(\sqrt{\frac{\pi e}{2}p^{2\rho}}\right),
\end{equation}
where $\rho$ is the code rate, $n$ is the lattice dimension and $p$ is the alphabet size.
Thus, the decoding of Construction A lattices  reaches to an error floor  which is caused by the uncoded layer. The authors of \cite{19} ensure the occurrence of this error floor in low error rates  by increasing the value of $p$ $(p = 11)$. Increasing the value of $p$ and using non-binary  LDPC codes as underlying code of Construction A lattices improves the error performance, but the penalty is increasing the complexity of encoding and decoding.
Using high rate ($\rho >0.83$) binary LDPC codes as underlying codes of the presented structure in \cite{sloane} and \cite[\S 20.5]{2} helps us to decrease the decoding complexity and avoid this error floor in symbols error rates more than $10^{-6}$ for $\textrm{VNR}=1$dB.

\subsection{Encoding of  QC-LDPC lattices}\label{encod2}
Following the suggested method  in \cite{sloane} and \cite[\S 20.5]{2}, the encoding  of QC-LDPC lattices can be performed using the following steps. First, convert the components of the codewords of $[n, k]$~binary code $\mathcal{C}$ into $\pm 1$ (convert $0$ to $-1$ and $1$ to $1$) \cite[\S 20.5]{2}, which produces a set $\Lambda(\mathcal{C})$ consisting of the
vectors of the form
\begin{equation}\label{newlattice}
  \mathbf{c}+4\mathbf{z}, \quad \mathbf{c}\in \mathcal{C},\,\, \mathbf{z}\in \mathbb{Z}^n.
\end{equation}
The set of the points in (\ref{newlattice})  strictly speaking is not a lattice, but the translate of a lattice by the vector $(-1, -1,\ldots , -1)$.
However, we can show that $\Lambda(\mathcal{C})$ is closed under following addition. In fact, for any $\bm{\lambda}_1,\bm{\lambda}_2\in\Lambda(\mathcal{C})$,  we have
\begin{equation}\label{sum}
  \bm{\lambda}_1\oplus\bm{\lambda}_2\triangleq\bm{\lambda}_1+\bm{\lambda}_2+(1,\ldots ,1)\in\Lambda(\mathcal{C}).
\end{equation}
Then, the encoding of an integer row vector $\mathbf{u}\in \mathbb{Z}^n$ is
\begin{equation}\label{encoding}
  \mathcal{E}(\mathbf{u})=2\mathbf{u}\mathbf{G}_{\Lambda}-(1,\ldots ,1),
\end{equation}
where $\mathcal{E}$ is the encoding function and $\mathbf{G}_{\Lambda}$ can be obtained based on Proposition~\ref{prop2} or  Theorem~\ref{prop3}. From  (\ref{VNR}) and (\ref{eq12}), the definition of VNR for this lattice is
\begin{equation}\label{SNR}
  \textrm{VNR}=\frac{4^{(2n-k)/n}}{2\pi e \sigma^2}.
\end{equation}
If $\mathbf{G}_{\Lambda}$ is of the form~(\ref{lattice_gen_main}), then from (\ref{VNR}) and (\ref{det_QC}) we have
\begin{equation}\label{SNR}
  \textrm{VNR}=\frac{4^{(tb+r)/tb}}{2\pi e \sigma^2}.
\end{equation}
\subsection{Encoding Complexity}

The complexity of an algorithm is a function describing the efficiency of the algorithm in terms of the amount of data the algorithm must process and there are different parameters for the domain and range of this function. Time and space complexity are different aspects for calculating the efficiency of an algorithm. The \emph{time complexity} of an algorithm quantifies the amount of time taken by an algorithm to run as a function of the input size. On the other hand, the \emph{space complexity} is a function describing the amount of memory (space) an algorithm takes in terms of the input size. In many cases,  we consider the extra memory needed, not counting the memory needed to store the input itself. There is often a time-space-tradeoff involved in a problem, that is, it cannot be solved with few computing time and low memory consumption at the same time. We have to make a compromise and to exchange computing time for memory consumption or vice versa. A good algorithm allows us to make this tradeoff between the number of steps (time complexity) and storage locations (space complexity).

In this subsection, we consider the encoding complexity of  QC-LDPC lattices. Using QC-LDPC codes instead of random LDPC codes as underlying codes of Construction A lattices, helps us to reduce the encoding complexity, which is essentially quadratic in the block length, into the practical values that are proportional to the block length. Indeed, using QC-LDPC lattices admits an encoding algorithm which has complexity that is linear in the dimension of the lattice $n$. This linearity is in both time and space domains.   We generalize the encoder circuit
of \cite{17} such that it can be used for encoding of QC-LDPC lattices. Then, we discuss the complexity of this circuit. The implemented encoder of \cite{17} offers a wide range of tradeoffs between
encoding speed and the space complexity of encoding  for QC-LDPC codes. Based on the proposed method in \cite{17}, encoding of QC-LDPC codes can be formed with \emph{shift-register-adder-accumulator} (SRAA) circuits.
The results of~\cite{17} show that for high-speed encoding of QC-LDPC codes, the complexity of the two-stage encoding is linearly proportional to the code length $n=tb$. For encoding of an integer vector $\mathbf{u}$, we partition it into two parts $\mathbf{u}_1$ and $\mathbf{u}_2$ of lengths $(tb-r)$ and $r$, respectively. Based on (\ref{lattice_gen_main}), we have
\begin{equation}\label{enc_part}
  \mathbf{u}\mathbf{G}_{\Lambda}=\mathbf{u}_1\mathbf{G}_{qc}^{*}+\mathbf{u}_2\mathbf{R}.
\end{equation}

\noindent The multiplication $\mathbf{u}_2\mathbf{R}$ can be done by concatenating one zero bit after least significant bit of components of $\mathbf{u}_2$ and then adding the $j^{th}$ component of the obtained vector by~$i_j^{th}$ component of  $\mathbf{u}_1\mathbf{G}_{qc}^{*}$, for $1\leq j\leq r$, where $i_j$ is defined in the proof of Theorem~\ref{prop3} (see Appendix~\ref{appB}). The computation of $\mathbf{u}_1\mathbf{G}_{qc}^{*}$ can be accomplished by changing each one of the encoder circuits of \cite{17} as follows. Assume that the components of information vector $\mathbf{u}$ are restricted to  the finite set of integers $\left\{-L,-L+1,\ldots ,L-1\right\}$, for $L\in\mathbb{Z}$. Let $w_c$ be the maximum column degree of $\mathbf{G}_{qc}^{*}$. Thus, the required number of bits for computing each component  of $\mathbf{u}\mathbf{G}_{\Lambda}$ is $N_b=\log_2\left(L\right)+w_c$. We should replace the XOR gates in encoder circuit of \cite{17} with $N_b$ bits full-adders and each AND gate with  $N_b$ AND gates. Since each $N_b$ bits full-adder contains a fixed number of AND-XOR gates,  the linear complexity of encoding by \cite{17} implies that the encoding of  QC-LDPC lattices can be done with linear complexity in the dimension of the lattice $n=tb$. Let $N_a$, $N_x$ and $N_o$ be the numbers of AND, XOR and OR gates, respectively, in each $N_b$ bits full-adder. \tablename~\ref{table0} gives the speeds and complexities of various encoding circuits of  QC-LDPC lattices. Note that the clock rate of these generalized circuites is lower than the clock rate of  their corresponding binary case in \cite{17}. This is a natural penalty for increasing the number of bits per each input symbol.

In order to make a comparison between regular encoding of QC-LDPC lattices and the proposed encoding methods in this paper, we present the encoding complexity by using the proposed generator matrix in  Proposition \ref{prop2}. Without lose of generality, let $\mathbf{T}=\mathbf{I}_n$ and the generator matrix of the considered QC-LDPC lattice $\Lambda$ be of the form given in (\ref{eq12}). For encoding an integer vector $\mathbf{u}$, we partition it into two parts $\mathbf{u}_1$ and $\mathbf{u}_2$ of lengths $k=cb$ and $n-k=(t-c)b$, respectively. Based on (\ref{eq12}), the encoded vector is $\boldsymbol{\lambda}=(\mathbf{u}_1,\mathbf{u}_1\mathbf{P}+2\mathbf{u}_2)$. Similar to the above, we only consider the complexity of computing $\boldsymbol{\lambda}_2=\mathbf{u}_1\mathbf{P}$. Computing $\boldsymbol{\lambda}_2$ needs $c(t-c)b^2$ multiplication and $(cb-1)(t-c)b$ addition. In this method, we need to store $\mathbf{P}$ entirely that needs $c(t-c)b^2$ flip-flops, because $\mathbf{P}$ has no specified structure. However, in the proposed encoders above, we only store $c(t-c)$ circulant generators \cite{17}. Similar to the above, define $N_b'=\log_2\left(L\right)+w_c'$, where $w_c'$ is the maximum column degree of $\mathbf{P}$. The speed and complexity of this encoding method is presented in \tablename~\ref{table0}. If all the symbols of $\boldsymbol{\lambda}_2$  are generated in parallel at the same
time, a circuit that completes encoding in  $1$ clock cycle can be implemented. In this case, $c(t-c)b^2$ registers
are needed. In this way, the encoding is completed in $1$ clock cycle. This implementation
requires a total of $c(t-c)b^2+cbN_b'$ flip-flops, $c((t-c)b-1)N_x$ XOR gates, $c((t-c)b-1)(N_a+N_b')$ AND gates and $c((t-c)b-1)N_o$ OR gates. This encoding is very fast but its implementation incurs intensive space complexity which is in contrast with  time-space-tradeoff.
\small
\begin{table}[h]
\caption{Comparison of different encoding schemes of  QC-LDPC lattices. }
\renewcommand{\arraystretch}{1}
\centering
\begin{tabular}{|K{2cm}||K{2cm}||K{2.5cm}||K{2.5cm}||K{2.5cm}||K{2.5cm}|}
\hhline{-||-||-||-||-||-}
Encoding   & Encoding   speed    & Flip-flops &Two input  &Two input  &Two input \\
scheme     & (Clock cycles)       &            & XOR gates &AND gates &OR gates\\
\hhline{=::=::=::=::=::=}
 SRAA (Serial Encoder) & $(t-c)b$   & $cb(N_b+1)$         &   $cbN_x$       & $cbN_b+cbN_a$              &$cbN_o$ \\
\hhline{-||-||-||-||-||-}
 SRAA (Parallel Encoder) & $cb$     & $(t-c)bN_b$      & $((t-c)b-1)N_x$   &$(t-c)bN_b+((t-c)b-1)N_a$ & $((t-c)b-1)N_o$\\
\hhline{-||-||-||-||-||-}
 Two-stage Encoder &   $b$          & $tbN_b$          & $O(N_xc^2b)$    &$O(N_ac^2b)$                &   $O(N_oc^2b)$   \\
 \hhline{-||-||-||-||-||-}
 Regular Encoder &  $1$ & $cb((t-c)b+N_b')$ & $c((t-c)b-1)N_x$ & $c((t-c)b-1)(N_a+N_b')$ &  $c((t-c)b-1)N_o$
 \\
  \hhline{-||-||-||-||-||-}
 \end{tabular}
\label{table0}
\end{table}
\normalsize
\subsection{Decoding of  QC-LDPC lattices}\label{decod2}
In this section, we propose two different decoding approaches for $\Lambda(\mathcal{C})=2\Lambda-(1,\ldots,1)$, where $\Lambda$ is a QC-LDPC lattice. The first one  is based on the proposed algorithm in \cite[\S 20.5]{2}. As a second method, we propose a new decoder  for QC-LDPC lattices based on SPA of LDPC codes, that has lower implementation complexity and lower memory requirements comparing to the first decoder. These decoding approaches are described in the rest of this section.
\subsubsection{Combination of SPA and Conway-Sloane's decoding method (CS-SPA) }
For decoding of $\Lambda(\mathcal{C})$, we plug in the SPA as a soft decoder into the Conway-Sloane decoding algorithm. The following lemma appeared in \cite{2}:
\begin{lemma}\label{lemm1}
Suppose $\mathbf{x} =(x_1,\ldots , x_n )$ lies in the cube $-1 \leq x_i \leq 1$, for $1\leq i\leq n$. Then, no point of $\Lambda(\mathcal{C})$ is closer to $\mathbf{x}$ than the closest codeword of $\mathcal{C}$.
\end{lemma}

To find the closest point of $\Lambda(\mathcal{C})$ to a given point $\mathbf{x}\in \mathbb{R}^n$, perform the steps given in \cite[\S 20.5, page 450]{2}. The implementation of the above algorithm for  QC-LDPC lattices is given in the sequel.
Let $\mathbf{x}=\mathbf{c}+4\mathbf{z}$ be the transmitted lattice vector as in (\ref{newlattice}) and  $\mathbf{y}$ be the received  vector from AWGN channel, therefore we have
\begin{equation}\label{AWGN_output1}
\mathbf{y}=\mathbf{c}+4\mathbf{z}+\mathbf{n},
\end{equation}
where $\mathbf{c}\in\mathcal{C}$  and $\mathcal{C}$ is a  QC-LDPC code with $\pm1$ components, $\mathbf{z}\in \mathbb{Z}^n$ and $\mathbf{n}\sim \mathcal{N}(0,\sigma^2)$. In first step, we decode $\mathbf{z}$ and the next step we find $\mathbf{c}$. Define $\hat{\mathbf{z}}$, the estimation of $\mathbf{z}$, as follows
\begin{equation}\label{AWGN_output1}
\hat{\mathbf{z}}=\left\lfloor \frac{\mathbf{y}-(1,\ldots,1)}{4}\right\rceil.
\end{equation}

Now define $a_i=y_i-4\hat{z}_i$, for $1\leq i\leq n$, and $S=\left\{1 \leq i\leq n\,\,|\,\, a_i>1\right\}$. Put
\begin{equation}\label{input_decoder}
\hat{a}_i=\left\{ \begin{array}{l}
                           2-a_i, \quad i\in S,\\
                            a_i, \quad \textrm{otherwise}.
                         \end{array}
\right.
\end{equation}

Sum-product algorithm (SPA) is a soft decision message-passing algorithm. For the sum-product decoder, the extrinsic information passed between nodes is also given as probabilities rather than hard decisions. Our proposed algorithms are similar to the SPA for LDPC codes in message passing structure \cite{sara}, but the input of our decoding algorithms are different from the SPA of LDPC codes.  The aim of SPA is computing the \emph{a posteriori probability} (APP)  for each codeword bit and to select the decoded value for each bit as the value with the maximum a posteriori probability (MAP). The SPA iteratively computes an approximation of the MAP  value for each code bit. The inputs are the log likelihood ratios (LLR) for the a priori message probabilities from each channel. Thus, we need to define log likelihood ratio for QC-LDPC lattices.
Define the $i^{th}$ LLR value as follows
\begin{equation}\label{LLR2}
\gamma_i =\frac{\left(\hat{a}_i+1\right)^2}{2\sigma^2}-\frac{\left(\hat{a}_i-1\right)^2}{2\sigma^2}.
\end{equation}
Input the LLR vector $\boldsymbol{\gamma}=(\gamma_1,\ldots, \gamma_n)$ to the  SPA decoder of the LDPC codes and consider $\tilde{\mathbf{c}}$ as the output of this decoder. Convert $\tilde{\mathbf{c}}$ to $\pm 1$ notation and call the obtained vector $\tilde{\mathbf{c}}'$. Define
\begin{equation}\label{input_decoder}
\hat{c}_i=\left\{ \begin{array}{l}
                           2-\tilde{c}'_i, \quad i\in S,\\
                            \tilde{c}'_i, \quad \textrm{otherwise}.
                         \end{array}
\right.
\end{equation}
Then, $\hat{\mathbf{x}}=\hat{\mathbf{c}}+4\hat{\mathbf{z}}$ is the decoded lattice vector.

\subsubsection{SPA of QC-LDPC lattices}
In this subsection, we introduce another decoding method for QC-LDPC lattices to decrease the decoding complexity. An application of the proposed decoding method in this subsection is also considered in a cooperative transmission framework \cite{IWCIT2015}.
Let $\mathbf{y}$ be as in~(\ref{AWGN_output1}).
In contrast to CS-SPA, first  we decode $\mathbf{c}$ and  next we find $\mathbf{z}$. This modification removes the considered memory for saving $S$ in CS-SPA. In this method, unlike the CS-SPA method that needs some pre-computations to estimate the LLR values, we estimate the LLR values directly from the received vector.    Define the $i^{th}$ LLR value as follows
\begin{eqnarray}\label{LLR}
\gamma_i &=&\log \left(\frac{\textrm{Pr}\left\{c_i=-1|y_i\right\}}{\textrm{Pr}\left\{c_i=+1|y_i\right\}}\right)\\
&\triangleq&2\left(\frac{ (\frac{y_i-1}{4}-\lfloor(\frac{y_i-1}{4})\rceil  )^2-  (\frac{y_i+1}{4}-\lfloor(\frac{y_i+1}{4})\rceil )^2}{\sigma^2}\right), \nonumber
\end{eqnarray}
where $\lfloor x\rceil$ is the nearest  integer to $x$. Input the LLR vector $\boldsymbol{\gamma}=(\gamma_1,\ldots, \gamma_n)$ to SPA decoder of LDPC codes and consider $\hat{\mathbf{c}}$ as the output of this decoder. Convert $\hat{\mathbf{c}}$ to $\pm 1$ notation and call the obtained vector $\hat{\mathbf{c}}'$. Estimate $\hat{\mathbf{z}}$ as follows
\begin{equation}\label{z_hat}
\hat{\mathbf{z}}=\left\lfloor\frac{\mathbf{y}}{4}-\frac{\hat{\mathbf{c}}'}{4}\right\rceil.
\end{equation}
Then, $\hat{\mathbf{x}}=\hat{\mathbf{c}}'+4\hat{\mathbf{z}}$ is the final decoded lattice vector. Decoding error happens when $\hat{\mathbf{x}} \neq \mathbf{x}$.
\subsection{Decoding complexity}
In this subsection we compare the decoding complexity of  QC-LDPC lattices with the decoding complexity of other well-known lattices that can be decoded with linear complexity in the dimension of lattice. Two families that we have considered are LDA lattices \cite{19} and LDLCs  \cite{LDLC}. The decoding algorithm of LDLCs with linear computational complexity first proposed in~\cite{LDLC} which has complexity $O(n\cdot d\cdot t\cdot \frac{1}{\Delta}\cdot \log_2(\frac{1}{\Delta}))$, where $\Delta$ is the resolution and its typical value is $1/256$, $n$ is the dimension of lattice, $t$ is the number of iterations and $d$ is the average code degree. Then, in~\cite{LDLC2}, a new algorithm proposed with lower complexity $O(n\cdot d\cdot t\cdot K\cdot M^3)$ compared to the one presented in \cite{LDLC}, where $K$ is the number of replications, $n$, $t$  and $d$ are similar to above. Proposed typical value for $K$ is $3$ and for $M$ is $2$ or $6$. The decoding complexity of LDA lattices  is $O(n\cdot d\cdot t\cdot p\cdot\log_2(p))$, where $p$ is the characteristic of the finite field that the underlying code is coming from \cite{19,LDAnew}. The least proposed value of $p$ is  $11$. The decoding  complexity of the both proposed algorithms in this paper, i.e. SPA and CS-SPA algorithms, are only $O(n\cdot d\cdot t)$,
because in each iteration  of them, $d$ multiplications per bit-node (in average) is required.
Thus, they have significantly lower complexity in comparison to the decoding algorithms of LDA lattices and LDLCs . In Section~\ref{Simulations}, we see that SPA and CS-SPA have the same performance. If we consider the implementation concerns, SPA is better than CS-SPA, because the implementation of CS-SPA needs  to save the indices in $S$ in each coming block of data which increases the memory requirement of CS-SPA in comparison to SPA.
 \section{Shaping methods of  QC-LDPC lattices}\label{shaping_sec}
In practical channels there exists a power constraint which is needed to be  fulfilled. This entails selecting a finite set of lattice points with bounded norms. In theoretical approaches, the coding lattice is intersected with a spherical shaping region to produce an efficient, power-constrained lattice code. However, spherical shaping has high computational complexity both for encoding and decoding.  In \cite{16}, several efficient and practical shaping algorithms proposed for LDLCs. In this section, we employ generator matrix of QC-LDPC lattices in conjunction with nested lattice shaping method to obtain QC-LDPC lattice codes. Another way of generating a QC-LDPC lattice code is to employ hypercube shaping algorithm given in \cite{16}.
\subsection{Nested lattice shaping method}
Nested lattice shaping has been proposed in \cite{0},  where the shaping domain of a lattice code is chosen as the Voronoi region of a different, coarse lattice, usually chosen as a scaled version of the coding lattice.
Let $\Lambda_s$ denotes the
scaled version of $\Lambda$ by $M$. A generator matrix $\mathbf{G}_s$ of $\Lambda_s$
can be derived by means of the generator matrix $\mathbf{G}$ of $\Lambda$ as
$\mathbf{G}_s=M\mathbf{G}$.
Let us assume that $\mathcal{V}$ and $\mathcal{V}_M$ represent the Voronoi cells of $\Lambda$ and
$\Lambda_s$ respectively.
Let $\Lambda$ be an $n$-dimensional lattice,
the code to be constructed, which is
called \emph{Voronoi code}, consists of all vectors $\mathbf{x}$
in $\Lambda\cap\mathcal{V}_M$. It contains $N=M^n$ codewords and has rate
$\log_2M$ bits per two dimensions. The bit labeling process can be performed easily.
Suppose that we want to encode $\mathbf{b}=(b_1,\ldots,b_n)$, $b_i\in\{0,1,\ldots,M-1\}$ for $1\leq i\leq n$.
Then, $\mathbf{x}_{\mathbf {b}}=\mathbf{b}\mathbf{G}-Q_{\Lambda_s}(\mathbf{b}\mathbf{G})$ and transmitted vector is $\mathcal{E}(\mathbf{x}_{\mathbf {b}})=2\mathbf{x}_{\mathbf {b}}-(1,\ldots,1)$,
where $Q_{\Lambda_s}$ is a quantizer for $\Lambda_s$.
The average power of $\Lambda\cap\mathcal{V}_M$ is estimated by means of the continues approximation~\cite{erez}.
The contribution of $\Lambda$ and $\mathcal{V}_M$ to the average power
can be separated as $P_{av}=G(\mathcal{V}_M)\det(\mathcal{V}_M)^{2/n}$,
where,
$$G(\mathcal{V}_M)=\frac{\int_{\mathcal{V}_M}\|\mathbf{x}\|^2d\mathbf{x}}{n\det(\mathcal{V}_M)^{1+2/n}}$$
is the normalized second moment of $\mathcal{V}_M$.
Note that $\det(\mathcal{V}_M)=M^n\det(\Lambda)$ and $\det(\mathcal{V}_M)^{2/n}$ depends only on
lattice $\Lambda$~\cite{tarokh, 3}.

The challenging part of this method is finding the closest point of coarse lattice to a specified point of fine lattice.  For
shaping applications it is not crucial to find the exact nearest lattice point, as the result will only be a slight penalty in signal power. The authors of \cite{16,3} have used LDLC decoder as a suboptimal quantizer.
The better the quantizer,  the better the shaping gain. The process of calculating the nested lattice shaping has been introduced in the sequel, briefly. As mentioned above, the hard and critical part of this process is the operation of quantizer, which is equivalent to solving the well-known \emph{Integer Least Squares} (ILS) problem.
To solve the ILS problem we use MILES \cite{m15}, which is an optimal quantizer, to obtain a better estimation of the shaping gain.

As mentioned in \cite{tarokh}, evaluation of the normalized second moment $G(\mathcal{B})$ is difficult, but it can be estimated by Monte Carlo integration. Based on the proposed encoding for 1-level LDPC and QC-LDPC lattices in this paper, coding and shaping lattices are $2\Lambda_c$ and $2\Lambda_s$, respectively. Let $\mathbf{x}_1 ,\ldots,\mathbf{x}_N$ be $N$ points uniformly distributed over $\mathcal{B}$. Then
\begin{eqnarray}
  \int_{\mathcal{B}}\|\mathbf{x}\|^2 d\mathbf{x} \approx \frac{V(\mathcal{B})}{N} \sum_{i=1}^N \|\mathbf{x}_i\|^2,
\end{eqnarray}
where $V(\mathcal{B})=M^n \det(2\Lambda_c)$. Note that, the translation of any region will not change its volume.  Thus,
\begin{eqnarray}
  G(\mathcal{B}) \approx \frac{\sum_{i=1}^N \|\mathbf{x}_i\|^2}{\det(2\Lambda_c)^{2/n}NnM^2}.
\end{eqnarray}

We also generated very high-dimensional QC-LDPC lattices and lattice codes (i.e., dimensions above $100$) using both nested and hypercube shaping methods. However, we do not include them here as the typical behavior of QC-LDPC codes is not the same at asymptotic dimensions versus small $n$.
\section{Numerical Analysis of QC-LDPC lattices}\label{Simulations}
\subsection{Numerical results of nested lattice shaping}
The derived numerical results of nested lattice shaping gain and shaping loss is presented in \tablename~\ref{table1}. All of the results of \tablename~\ref{table1} are obtained by considering constellation size $4$.  In order to obtain the  exact shaping gain, we have used the optimal quantizer of \cite{m15}. This quantizer searches without restriction to find the exact closest vector.
 We have used random QC-LDPC codes of sizes $(40,20)$, $(50,25)$ and $(60,30)$ as underlying codes of  QC-LDPC lattices.
 \begin{table}[h]
\small
\caption{Estimated shaping gain of QC-LDPC lattices using nested lattice shaping method. }
\renewcommand{\arraystretch}{1.2}
\centering
\begin{tabular}{|c||c||c|}
\hhline{-||-||-}
Dimension of Lattice     & Shaping Gain (dB)  &Shaping Loss (dB) \\
\hhline{=::=::=}
  40 & 0.512 & 0.707\\
  50 & 0.577 & 0.687\\
  60 & 0.627 & 0.668\\
  \hhline{-||-||-}
 \end{tabular}
\label{table1}
\end{table}
\subsection{Error performance of QC-LDPC lattices}\label{LDPCLatticesareLDLC}
\begin{figure}[ht]
\centering
\includegraphics[width=12cm]{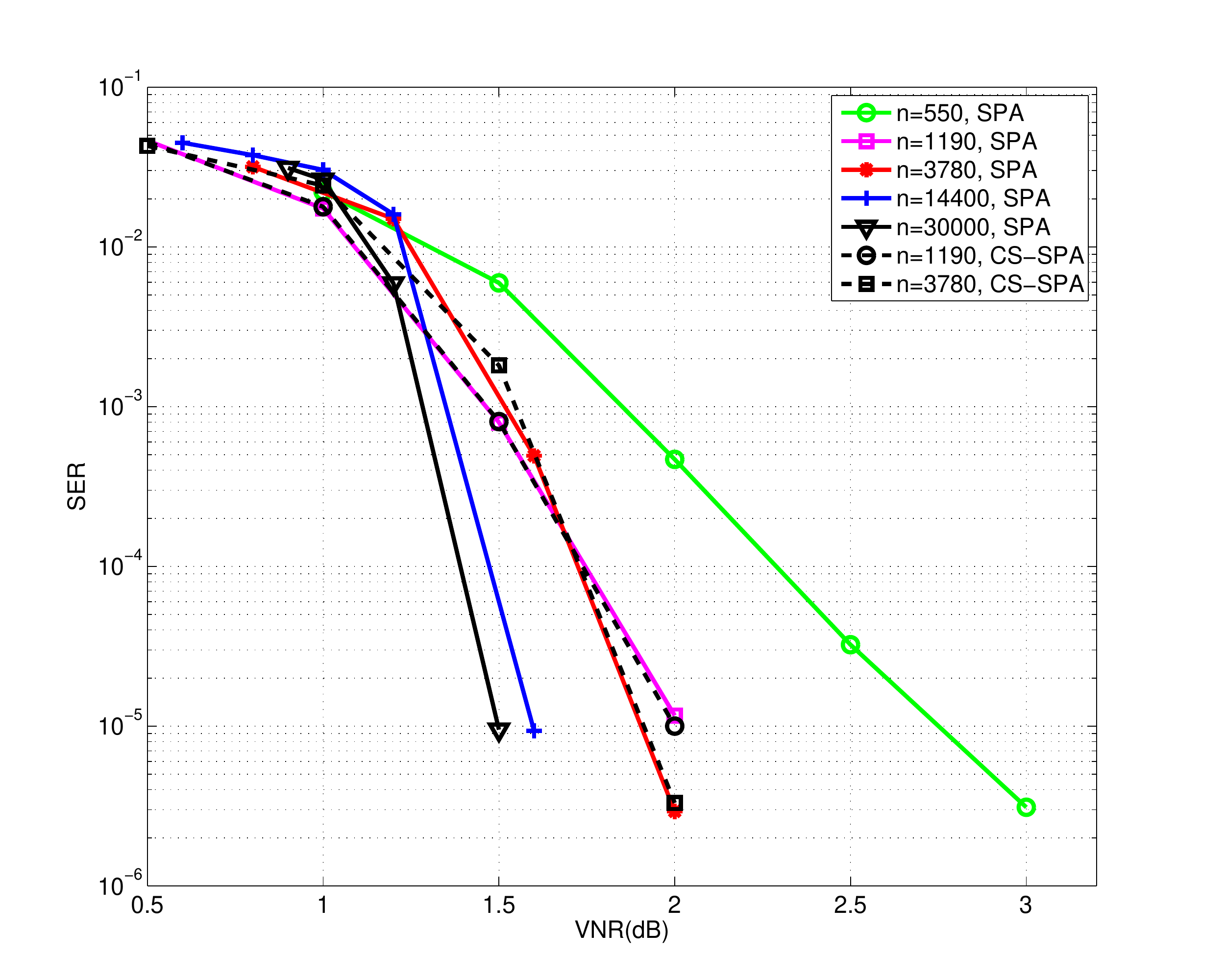}
\caption{Symbol error performance of QC-LDPC lattices with different dimensions.}
\label{G-SPA}
\end{figure}

The simulation results of the decoding performance of  QC-LDPC lattices  using SPA and CS-SPA are presented in \figurename{\ref{G-SPA}}.
We have used random girth $8$ QC-LDPC codes of sizes $(30000,25000)$, $(14400,12000)$, $(3780,3240)$, $(1190,935)$, and $(550,458)$ as underlying codes of  QC-LDPC lattices. The maximum number  of iterations in all of the simulations is $50$.   The   QC-LDPC lattices of sizes $n=1190$ and $n=30000$, at SER of $10^{-5}$, can work $2$dB and $1.5$dB  away from the capacity, respectively. We also compared the SER performance of  QC-LDPC lattices with dimensions $3780$ and $1190$ by using SPA and CS-SPA decoding methods. We observe that both of the algorithms have almost the same performance. LDA lattice \cite{19} of dimension $1000$ attains a SER of $10^{-5}$ at $1.35$dB from capacity.  The performance of LDLC lattice \cite{LDLC} of dimension $1000$ at a SER of $10^{-5}$ is at $1.7$dB from capacity.
Hence, the error performance of an LDA lattice and an LDLC of dimension $1000$ are $0.65$dB and $0.3$dB better than the error performance of a QC-LDPC lattice of dimension $1190$, respectively. However, the  decoding complexity of LDA lattices and LDLCs  are at least
$38$ and $24$ times more than the decoding complexity of  QC-LDPC lattices.
Indeed, to have a fair comparison in terms of complexity, one would use a QC-LDPC lattice of dimension 30000 with an LDA lattice of dimension $780$ and an LDLC of dimension $1250$ instead.
Thus, the simulations indicate that the proposed lattice codes come close to matching the performance of LDA lattices
and LDLCs, with significant savings in encoding and decoding complexity.
\section{Concluding remarks}\label{ConclusionandSuggestionforLaterResearch}
The  QC-LDPC lattices have been analysed.  These lattices are equivalent to the  Construction A lattices which are a lifting of a binary QC-LDPC code. The generator matrix of these lattices are obtained in such a way that they  can be encoded/decoded with lowest memory requirement and complexity. Experimental results show that if we consider equal decoding complexity,  they  have good error performance compared to their competitors such as LDLCs \cite{LDLC} and LDA lattices \cite{19}.
Unlike the LDA and LDLC lattices, QC-LDPC lattices have linear encoding  complexity. Decoding complexity of  QC-LDPC lattices is also significantly lower than  both  LDA lattices and LDLCs, which makes them a good choice for practical implementation.
The value of shaping gain of these lattices shows that they can be considered as one of the best finite constellations.  In a nutshell, one can extract good lattice codes from  QC-LDPC lattices which are appropriate for both Rayleigh fading and AWGN channels~\cite{viterbo}.

\appendices
\section{Basic properties of matrices over $\mathbb{Z}$}\label{app0}
Here, we give the necessary definitions and results about the properties of matrices over $\mathbb{Z}$.
%
%
The set of all $m \times n$ matrices with entries from $\mathbb{Z}$ will be denoted by $\mathbb{Z}^{m\times n}$.
The set of invertible matrices in $\mathbb{Z}^{n\times n}$ is denoted by $\textrm{GL}_n(\mathbb{Z})$.  Every member of $\textrm{GL}_n(\mathbb{Z})$ is called a \emph{unimodular matrix} over $\mathbb{Z}^{m\times n}$. Two matrices $\mathbf{A},\mathbf{B} \in \mathbb{Z}^{m\times n}$ are  equivalent if there exist matrices $\mathbf{U}\in \textrm{GL}_m(\mathbb{Z})$ and $\mathbf{V} \in \textrm{GL}_n (\mathbb{Z})$ such that $\mathbf{UAV} = \mathbf{B}$.  A diagonal matrix $\mathbf{D} = \textrm{diag}(d_1 ,\ldots,d_r ) \in \mathbb{Z}^{m\times n}$ $(r = \min\left\{n,m\right\})$ is called a \emph{Smith normal form of} $\mathbf{A}$, if $\mathbf{D}$ is equivalent to $\mathbf{A}$ and $d_1 | d_2 | \cdots | d_r$. Matrix $\mathbf{B}$ is obtained by applying elementary operations on the rows and columns of a $\mathbf{A}$ each being accomplished by multiplying $\mathbf{A}$ on the left and right by  unimodular matrices $\mathbf{U}$ and $\mathbf{V}$, respectively.
It is known \cite{PID} that every matrix over $\mathbb{Z}$ has a Smith normal form whose diagonal entries are unique up to equivalence of associates. Hence, we have the following consequence.
\begin{Corollary}\label{cor1}
If  $\mathbf{A}\in \mathbb{Z}^{m\times n}$, then column rank and row rank of $\mathbf{A}$ are equal.
\end{Corollary}
\section{Proof of Theorem~\ref{prop3}}\label{appB}
To prove Theorem~\ref{prop3} we need the following lemmas.
\begin{lemma}\label{ind_pos_lemma}
Let $\mathbf{v}_1,\ldots,\mathbf{v}_n\in \mathbb{F}^m$ be column vectors where $\mathbb{F}$ is an arbitrary field. Consider $\mathbf{w}_i$ as a vector that we obtain from $\mathbf{v}_i$, for $1\leq i \leq n$, by choosing $r$ components in arbitrary  positions $i_1,\ldots , i_r$. If $\mathbf{w}_1,\ldots ,\mathbf{w}_n$ are linearly independent over  $\mathbb{F}$, then $\mathbf{v}_1,\ldots,\mathbf{v}_n$ will be linearly independent over $\mathbb{F}$.
\begin{IEEEproof}
The proof is trivial.
\end{IEEEproof}
\end{lemma}

Based on Lemma~\ref{ind_pos_lemma}, since first $d_i$ columns of $\mathbf{Q}_{i,i}$ are linearly independent over $\mathbb{F}_2$, for $1\leq i\leq l$, their corresponding columns in $\mathbf{G}_{qc}^{*}$ are also linearly independent over $\mathbb{F}_2$.
Due to the cyclic structure of circulants,  the last $\bar{d}_i$ columns of the
$i^{th}$ column of circulants in $\mathbf{G}_{qc}^{*}$ can be considered as linearly dependent columns. Thus, the first $d_1,d_2,\ldots d_l$ columns of the $1^{st}$, $2^{nd}$,$\cdots$, $l^{th}$  columns of circulants of $\mathbf{G}_{qc}^{*}$ are linearly independent columns over $\mathbb{F}_2$. Now, we show that these columns are also linearly independent over the bigger ring $\mathbb{Z}$.
\begin{lemma}\label{ind_Z2}
Let $\mathbf{v}_1,\ldots,\mathbf{v}_n\in \mathbb{F}_p^n$, where $p$ is a prime number. If $\mathbf{v}_1,\ldots,\mathbf{v}_n$ are linearly independent over $\mathbb{F}_p$, then they are linearly independent over $\mathbb{Z}$.
\begin{IEEEproof}
Let $c_1\mathbf{v}_1+\cdots +c_n\mathbf{v}_n=\mathbf{0} $  where $c_1,\ldots,c_n\in \mathbb{Z}$. This implies that $\bar{c}_1\mathbf{v}_1+\cdots +\bar{c}_n\mathbf{v}_n=\mathbf{0} $, where $\bar{c}_i=c_i\pmod{p}$, for $1\leq i \leq n$. Since $\mathbf{v}_1,\ldots,\mathbf{v}_n$ are linearly independent over $\mathbb{F}_p$, we must have $c_i=\alpha_i p^{\beta_i}$,  where $\alpha_i\in \mathbb{Z}$ and $\beta_i\in \mathbb{Z}^{+}$ and $\textrm{gcd}(\alpha_i,p)=1$, for $1\leq i\leq n$. Define $\beta=\beta_k=\underset{1\leq i\leq n}{\min} \beta_i$. Since $\mathbb{Z}$ has no zero divisor, we have
$$\alpha_1p^{\beta_1-\beta}\mathbf{v}_1+\cdots +\alpha_k\mathbf{v}_k+\cdots +\alpha_np^{\beta_n-\beta}\mathbf{v}_n=\mathbf{0},$$
which implies $c_1'\mathbf{v}_1+\cdots +c_n'\mathbf{v}_n=\mathbf{0} $, where $c_i'=p^{-\beta}c_i\pmod {p}$. This is a contradiction because $c_k'=\alpha_k$ and $\textrm{gcd}(\alpha_k,p)=1$.
\end{IEEEproof}
\end{lemma}

We know that the rows of $\mathbf{G}_{qc}^{*}$ together with the rows of the form $2\mathbf{e}_i$ for $1\leq i\leq n$,  generate every vector in $\Lambda$.
In any vector space $\mathcal{M}$ over  field $\mathbb{F}$, any generating subset of $\mathcal{M}$ contains a basis of $\mathcal{M}$. If $\mathbb{F}$ is not a field, this statement becomes completely false
\footnote{
For example consider $\mathcal{M}=\mathbb{Z}$ and $\mathbb{F}=\mathbb{Z}$. Then $\mathcal{M}$ has a basis $\mathcal{X}$, in fact there are just two possibilities: $\mathcal{X}=\left\{1\right\}$ or $\mathcal{X}=\left\{-1\right\}$. However, $\mathcal{X}=\left\{2,3\right\}$ is a generating set which does not contain a basis.}.
Now we are ready to give the proof of Theorem~\ref{prop3}.

\textbf{Proof of Theorem~\ref{prop3}:} It is clear that the rows of $\mathbf{G}_{qc}^{*}$ together with the rows of the form $2\mathbf{e}_i$ for $1\leq i\leq n$, will generate every vector in $\Lambda$.
Based on Lemma~\ref{ind_Z2}, since the rows of $\mathbf{G}_{qc}^{*}$ are linearly independent over $\mathbb{F}_2$, they are also linearly independent over $\mathbb{Z}$.
We consider the rows of  $\mathbf{G}_{qc}^{*}$ as first part of a generating set for $\Lambda$. For $i=1,\ldots,n$, it is clear that $2\mathbf{e}_i\in \Lambda$. Let $\mathbf{g}_1,\ldots , \mathbf{g}_{tb-r}$ be the rows of $\mathbf{G}_{qc}^{*}$.  We  find $1\leq i_1,\ldots , i_r\leq n$ such that $\left\{\mathbf{g}_1,\ldots ,\mathbf{g}_{tb-r},2\mathbf{e}_{i_1},\ldots, 2\mathbf{e}_{i_r}\right\}$ form a basis for $\Lambda$. The column and row rank of $\mathbf{G}_{qc}^{*}$ are equal  (Corollary~\ref{cor1}). Hence,  $\mathbf{G}_{qc}^{*}$ has exactly $r$ dependent columns over $\mathbb{Z}$. We claim that $i_1,\ldots,i_r$ are exactly the positions of these dependent columns. We  find these positions as follows. Define $\bar{d}_0=0$, then we have $\sum_{i=0}^l\bar{d}_i=r$. Thus, for each $1\leq k\leq r$ there exists $0\leq j\leq l-1$ such that $\bar{d}_j\leq k<\bar{d}_{j+1}$. Then
\begin{equation}\label{ik}
  i_k=(t+j-l)b+d_{j+1}+k-\bar{d}_j+1.
\end{equation}
We prove  that these vectors are linearly independent over $\mathbb{Z}$. Let $\mathbf{g}^{(j_1)},\ldots,\mathbf{g}^{(j_{tb-r})}$ be the independent columns of $\mathbf{G}_{qc}^{*}$.
Put $\mathbf{G}_1=\left[
                                                     \begin{array}{cc}
                                                       \mathbf{G}_{qc}^{*t} &
                                                       2\mathbf{e}_{i_1}^{t}
                                                     \end{array}
                                                   \right]^t$.
Now, consider the columns $i_1,j_1,j_2,\ldots , j_{tb-r}$ of $\mathbf{G}_1$ and call them $\mathbf{g}_1^{(i_1)},\mathbf{g}_1^{(j_1)},\ldots ,\mathbf{g}_1^{(j_{tb-r})}$, respectively. If
$$\beta_1\mathbf{g}_1^{(i_1)}+\beta_2\mathbf{g}_1^{(j_1)}+\cdots +\beta_{tb-r+1}\mathbf{g}_1^{(j_{tb-r})}=\mathbf{0},$$
then
$$\beta_1\left[
\begin{array}{l}
\mathbf{g}^{(i_1)}\\
2
\end{array}
\right]+\beta_2\left[
\begin{array}{l}
\mathbf{g}^{(j_1)}\\
0
\end{array}
\right]+\cdots +\beta_{tb-r+1}\left[
\begin{array}{l}
\mathbf{g}^{(tb-r)}\\
0
\end{array}
\right]=\mathbf{0}.$$
Thus $\beta_1=0$ and $\beta_2\mathbf{g}^{(j_1)}+\cdots +\beta_{tb-r+1}\mathbf{g}^{(j_{tb-r})}=\mathbf{0}$ that implies $\beta_2=\cdots =\beta_{tb-r}=0$. Therefore,  the column rank of $\mathbf{G}_1$ is $tb-r+1$ and consequently
the rows of $\mathbf{G}_1$ are linearly independent over $\mathbb{Z}$. By induction and considering $\mathbf{G}_1$ instead of $\mathbf{G}_{qc}^{*}$ we can prove the result. Hence, the considered vectors are linearly independent over $\mathbb{Z}$. It is enough to show that these vectors generate every point in lattice $\Lambda$. Indeed, we must show that vectors of the form $2\mathbf{e}_i$ where $i\neq i_1,\ldots,i_r$, will be generated by $\mathcal{B}=\left\{\mathbf{g}_1,\ldots,\mathbf{g}_{tb-r},2\mathbf{e}_{i_1},\ldots,2\mathbf{e}_{i_{r}}\right\}$. Put the members of $\mathcal{B}$ as rows of the matrix $\mathbf{G}'$ and call $\Lambda'$, the generated lattice by $\mathbf{G}'$.
First, we show that the determinant of lattice $\Lambda=\mathcal{C}+2\mathbb{Z}^n$ and $\Lambda'$ are both equal to $2^r$. It is clear that $\Lambda'$ is a sublattice of $\Lambda$ and both $\Lambda$ and $\Lambda'$ have the same rank $n$. Thus, if we show that $\det(\Lambda)=\det(\Lambda')$, then $\Lambda=\Lambda'$ and we obtain the desired result. Let $\mathcal{C}_{sys}$ with generator matrix $\mathbf{G}_{sys}=[\mathbf{I}_{tb-r}\,\, \mathbf{P}']$, be the systematic version of code $\mathcal{C}$. Then, the generator matrix of lattice $\Lambda''=\mathcal{C}_{sys}+2\mathbb{Z}^n$ has the following form
\begin{equation}\label{lambda_gen}
  \mathbf{G}''=\left[
                           \begin{array}{cc}
                             \mathbf{I}_{tb-r} & \mathbf{P}' \\
                             \mathbf{0}_{r\times tb-r} & 2\mathbf{I}_{r} \\
                           \end{array}
                         \right].
\end{equation}
The codes $\mathcal{C}_{sys}$ and $\mathcal{C}$ are equivalent, hence we can obtain codewords of $\mathcal{C}_{sys}$ by applying a fixed permutation $\pi$ on codewords of $\mathcal{C}$. Indeed, there is a fixed permutation $\sigma$ on $\left\{1,\ldots,n\right\}$ such that the map $\pi$
\begin{eqnarray*}
  \pi: \mathcal{C} &\longrightarrow& \mathcal{C}_{sys} \\
  (c_1,c_2,\ldots,c_n)&\longmapsto& (c_{\sigma(1)},c_{\sigma(2)},\ldots,c_{\sigma(n)}),
\end{eqnarray*}
is a bijection.
Let $\mathbf{c}''\in \mathcal{C}_{sys}$ and $\mathbf{c}''=\pi(\mathbf{c})$ for $\mathbf{c}\in \mathcal{C}$. For each $\mathbf{z}''\in \mathbb{Z}^n$, $\bm{\lambda}''=\mathbf{c}''+2\mathbf{z}''$ belongs to $\Lambda''$ and we can find $\mathbf{z}\in\mathbb{Z}^n$ such that $\mathbf{z}''=\pi(\mathbf{z})$. It is clear that
\begin{eqnarray*}
  \bm{\lambda} &=& \mathbf{c}+2\mathbf{z}=\pi^{-1}(\mathbf{c}'')+2\pi^{-1}(\mathbf{z}'') \\
  &=& \pi^{-1}(c+2z''),
\end{eqnarray*}
belongs to $\Lambda$. Indeed, for each $\bm{\lambda}\in \Lambda$, $\pi(\bm{\lambda})\in\Lambda''$. Thus, the generator matrix of $\Lambda$ can be obtained by multiplying $\mathbf{G}''$ from the right by permutation matrix $\textbf{T}$ of $\pi^{-1}$, i.e., $\mathbf{G}_{\Lambda}=\mathbf{G}''\textbf{T}$. Therefore, $\det(\Lambda)=\det(\Lambda'')=2^r$. It is enough to show that $\det(\mathbf{G}')=2^r$. The Generator matrix of $\Lambda'$ is of the form (\ref{lattice_gen_main}). The submatrix $\mathbf{Q}$ of $\mathbf{G}'$ is formed itself by $\mathbf{Q}_{i,j}$'s, for $1\leq i,j\leq l$. If $i=j$, then $\mathbf{Q}_{i,j}$ is a $d_i\times b$ matrix of the following $3$ forms. In all of the following matrices, ``$*$'' represents elements that can be $0$ or $1$. The big zeros denote that the specified parts by lines in the matrix are zero.
\newcommand\x{\ast}
\newcommand*{\bord}{\multicolumn{1}{|c}{}}
\newcommand*{\bors}{\multicolumn{1}{c|}{}}
\newcommand*{\borsd}{\multicolumn{1}{|c|}{}}
\begin{enumerate}
  \item If $\bar{d}_i=d_i$, then
\begin{equation}\label{Q_ii1}
 \mathbf{Q}_{i,i}=\left[
                    \begin{array}{ccc|ccc}
                    \cline{1-1}  1  & \bord & \textbf{\Large0} & \x & \x & \x   \\ \cline{2-2} \cline{4-4}
                      \x & 1 &\borsd &  \bors&  \x & \x \\\cline{3-3} \cline{5-5}
                      \x & \x & 1 & \textbf{\Large0} & \bors & \x \\ 
                    \end{array}
                  \right].
\end{equation}
\item If $\bar{d}_i<d_i$, then
\begin{equation}\label{Q_ii2}
 \mathbf{Q}_{i,i}=\left[
                    \begin{array}{cccc|ccc}
                      \cline{1-1} 1 & \bord &  & \textbf{\Large0} & \x & \x & \x \\ \cline{2-2} \cline{5-5}
                      \x & 1 & \bord &  & \bors & \x & \x \\ \cline{3-3} \cline{6-6}
                      \x & \x & 1 & \borsd & \textbf{\Large0} & \bors & \x  \\ \cline{4-6}
                      \x & \x & \x & 1 & \x & \x & \x  \\ 
                    \end{array}
                  \right].
\end{equation}
  \item If $\bar{d}_i>d_i$, then
\begin{equation}\label{Q_ii3}
\mathbf{Q}_{i,i}=  \left[
    \begin{array}{ccc|cccc}
      \cline{1-1} 1 & \bord & \textbf{\Large0}            & \x & \x & \x& \x \\ \cline{2-2} \cline{4-4}
      \x & 1 & \borsd  & \bors & \x & \x & \x \\\cline{3-3} \cline{5-5}
      \x & \x & 1            & \textbf{\Large0} & \bors & \x & \x \\
    \end{array}
  \right].
\end{equation}
\end{enumerate}

For $1\leq i,j\leq l$, if $i\neq j$, then $\mathbf{Q}_{i,j}$ has the following $3$ cases.
\begin{enumerate}
  \item If $d_i=d_j$ then
\begin{equation}\label{Q_ij1}
 \mathbf{Q}_{i,j}=\left[
                    \begin{array}{ccc|ccc}
                      \cline{1-1} 0 & \bord & \textbf{\Large0}& \x & \x& \x \\ \cline{2-2} \cline{4-4}
                      \x & 0 & \borsd & \bors & \x & \x \\ \cline{3-3} \cline{5-5}
                      \x & \x & 0 & \textbf{\Large0} & \bors & \x \\
                    \end{array}
                  \right].
\end{equation}
  \item If $d_i<d_j$ then
\begin{equation}\label{Q_ij2}
   \mathbf{Q}_{i,j}=\left[
                    \begin{array}{ccccc|ccccc}
                      \cline{1-1} 0 & \bord &  &  & \textbf{\Large0}    & \x  & \x  & \x  & \x  & \x  \\ \cline{2-2} \cline{6-6}
                      \x  & 0 & \bord  &  &      & \bors & \x  & \x  & \x  & \x  \\ \cline{3-3}  \cline{7-7}
                 \x  & \x  & 0 & \bord  &   &   &  \bors &  \x   & \x    & \x  \\ \cline{4-4}  \cline{8-8}
                      \x  & \x  & \x   & 0 &  \borsd    &  \textbf{\Large0} &   &  \bors  & \x     &\x \\ \cline{5-5}
                    \end{array}
                  \right],
\end{equation}
where the left part of $\mathbf{Q}_{i,j}$ is a $d_i\times d_j$ matrix.
  \item If $d_i>d_j$ then
\begin{equation}\label{Q_ij3}
  \mathbf{Q}_{i,j}= \left[
  \begin{array}{cccc|cccc}
   \cline{1-1} 0 & \bord & & \textbf{\Large0}& \x & \x & \x & \x \\ \cline{2-2} \cline{5-5}
   \x & 0 &\bord & & \bors & \x& \x &\x\\ \cline{3-3} \cline{6-6}
   \x& \x &0 & \borsd& & \bors &  \x & \x\\ \cline{4-4} \cline{7-7}
   \x& \x & \x &    0&  &  &  \bors &    \x\\ \cline{8-8}
   \x & \x & \x &    \x &  \textbf{\Large0}&  &   &    \\
  \end{array}
  \right].
\end{equation}
\end{enumerate}

By expanding the determinant of $\mathbf{G}'$ along the  first $t-l$ columns and then expanding it along the last $r$ rows, we obtain $\det(\mathbf{G}')=2^r\det (\mathbf{Q}')$, where $\mathbf{Q}'$ is a submatrix of $\mathbf{Q}$ that is obtained by removing the columns $i_1-t+l,\ldots,i_r-t+l$. $\mathbf{Q}'$ is an $(lb-r)\times (lb-r)$ matrix formed by $\mathbf{Q}_{i,j}'$'s, where $\mathbf{Q}_{i,j}'$ is the left part of $\mathbf{Q}_{i,j}$, for $1\leq i,j \leq l$. We show that $\det(\mathbf{Q}')=1$. First, we state a useful result about the determinant of block matrices \cite{Silve}. Let $\mathbf{M}$ be the following matrix
\begin{equation}\label{silvester_mat}
  \mathbf{M}=\left[
      \begin{array}{cc}
        \mathbf{A} & \mathbf{B} \\
        \mathbf{C} & \mathbf{D} \\
      \end{array}
    \right],
\end{equation}
where $\mathbf{A}, \mathbf{B}, \mathbf{C}$, and $\mathbf{D}$ are $k \times k$, $k \times (n- k)$, $(n -k) \times k$, and $(n-k) \times (n- k)$ matrices, respectively. Then
\begin{equation}\label{silvester_identity}
  \det(\mathbf{M})=\det(\mathbf{D})\det\left(\mathbf{A}-\mathbf{B}\mathbf{D}^{-1}\mathbf{C}\right).
\end{equation}

The matrix $\mathbf{A}-\mathbf{B}\mathbf{D}^{-1}\mathbf{C}$ is called the \emph{Sch\"{u}r
complement} with respect to $\mathbf{D}$ \cite{schur}. We prove by induction on $l$ (the number of blocks in each row and column of $\mathbf{Q}'$) that $\det(\mathbf{Q}')=1$.
If $l=1$ then
\begin{equation}\label{Q_ii1}
 \mathbf{Q}'=\left[
                    \begin{array}{ccc}
                    \cline{1-1}  1  & \bord & \textbf{\Large0}   \\ \cline{2-2}
                      \x & 1 &\bord   \\ \cline{3-3}
                      \x & \x & 1  \\ 
                    \end{array}
                  \right],
\end{equation}
which has determinant $1$. Assume that the result is true for $l-1$  and $\mathbf{Q}'$ contains $l$ blocks in each row or column.  $\mathbf{Q}'$ has the following form
\begin{equation}\label{Q'}
  \mathbf{Q}'=\left[
                \begin{array}{c|c}
                   \mathbf{Q}'(1:l-1,1:l-1) & \begin{array}{c}
                                         \mathbf{Q}'_{1,l} \\
                                         \mathbf{Q}'_{2,l} \\
                                         \vdots \\
                                         \mathbf{Q}'_{l-1,l}
                                       \end{array}
                    \\
                    \hline
                  \begin{array}{cccc}
                    \mathbf{Q}'_{l,1}  & \mathbf{Q}'_{l,2} & \cdots & \mathbf{Q}'_{l,l-1}
                  \end{array}
                   & \mathbf{Q}'_{l,l}  \\
                \end{array}
              \right].
\end{equation}

Consider
\begin{eqnarray*}
  \mathbf{A} &=& \mathbf{Q}'(1:l-1,1:l-1), \\
  \mathbf{B} &=& \left[ \begin{array}{cccc}
                                         \mathbf{Q}_{1,l}^{'t} &
                                         \mathbf{Q}_{2,l}^{'t}&
                                         \cdots &
                                         \mathbf{Q}_{l-1,l}^{'t}
                                       \end{array} \right]^t, \\
  \mathbf{C} &=& \left[\begin{array}{cccc}
                    \mathbf{Q}'_{l,1}  & \mathbf{Q}'_{l,2} & \cdots & \mathbf{Q}'_{l,l-1}
                  \end{array}\right], \\
  \mathbf{D} &=& \mathbf{Q}'_{l,l},
\end{eqnarray*}
and apply equation (\ref{silvester_identity}). It is also clear that $\det(\mathbf{D})=\det(\mathbf{Q}'_{l,l})=1$.
We also have
\begin{eqnarray*}
\mathbf{B}\mathbf{D}^{-1}\mathbf{C}\!\!\!\!&=&\!\!\!\!\left[\begin{array}{c}
                                         \mathbf{Q}'_{1,l} \\
                                         \mathbf{Q}'_{2,l} \\
                                         \vdots \\
                                         \mathbf{Q}'_{l-1,l}
                                       \end{array}\right] \mathbf{Q}_{l,l}^{'-1}  \left[\begin{array}{cccc}
                    \mathbf{Q}'_{l,1}  & \mathbf{Q}'_{l,2} & \cdots & \mathbf{Q}'_{l,l-1}
                  \end{array}\right]\\
                 \!\!\!\! &=&\!\!\!\!\left[
                       \begin{array}{ccc}
                          \mathbf{Q}'_{1,l}\mathbf{Q}_{l,l}^{'-1}\mathbf{Q}'_{l,1}  & \cdots & \mathbf{Q}'_{1,l}\mathbf{Q}_{l,l}^{'-1}\mathbf{Q}'_{l,l-1}  \\
 \mathbf{Q}'_{2,l}\mathbf{Q}_{l,l}^{'-1}\mathbf{Q}'_{l,1}   & \cdots & \mathbf{Q}'_{2,l}\mathbf{Q}_{l,l}^{'-1}\mathbf{Q}'_{l,l-1}  \\
                         \vdots & \ddots & \vdots \\
 \mathbf{Q}'_{l-1,l}\mathbf{Q}_{l,l}^{'-1}\mathbf{Q}'_{l,1}   & \cdots & \mathbf{Q}'_{l-1,l}\mathbf{Q}_{l,l}^{'-1}\mathbf{Q}'_{l,l-1}  \\
                       \end{array}
                     \right]\\
                    \!\!\!\! &=&\!\!\!\!\left[
                          \begin{array}{cccc}
                            \mathbf{Q}''_{1,1} & \mathbf{Q}''_{1,2} & \cdots & \mathbf{Q}''_{1,l-1} \\
                            \mathbf{Q}''_{2,1} & \mathbf{Q}''_{2,2} & \cdots & \mathbf{Q}''_{2,l-1} \\
                            \vdots & \vdots & \ddots & \vdots \\
                            \mathbf{Q}''_{l-1,1} & \mathbf{Q}''_{l-1,2} & \cdots & \mathbf{Q}''_{l-1,l-1} \\
                          \end{array}
                        \right].
\end{eqnarray*}

For $1\leq i,j\leq l-1$, if $\mathbf{Q}'_{i,j}$ is the left part of $\mathbf{Q}_{i,j}$ in (\ref{Q_ij1}), then $\mathbf{Q}'_{j,i}$ will also be of the same form, if $\mathbf{Q}'_{i,j}$ is the left part of $\mathbf{Q}_{i,j}$ in (\ref{Q_ij2}), then $\mathbf{Q}'_{j,i}$ will  be the  left part of $\mathbf{Q}_{j,i}$ which is  of the form (\ref{Q_ij3}) and vice versa. Since,  all of  $\mathbf{Q}'_{i,j}$'s are rectangular (or square) lower triangular   matrices, it can be easily checked that for $1\leq i,j\leq l-1$, $\mathbf{Q}''_{i,j}=\mathbf{Q}'_{i,l}\mathbf{Q}_{l,l}^{'-1}\mathbf{Q}'_{l,j} $ is a lower triangular matrix of size $d_i\times d_j$ (which is the size of $\mathbf{Q}'_{i,j}$), where all elements on the main diagonal are zero. Note that $\mathbf{Q}_{l,l}^{'-1}$ is also a lower triangular matrix.  By computing $\mathbf{A}-\mathbf{B}\mathbf{D}^{-1}\mathbf{C}$, we get a matrix $\hat{\mathbf{Q}}$ that is formed by $(l-1)\times (l-1)$ blocks and all of the blocks are lower triangular matrices. Main diagonal blocks of  $\hat{\mathbf{Q}}$ are square lower triangular matrices with $1$ on their main diagonal and other blocks of $\hat{\mathbf{Q}}$ are lower triangular matrices with $0$ on their main diagonal. Hence, $\hat{\mathbf{Q}}$ fulfills in the induction hypothesize. Thus, $\det\left(\hat{\mathbf{Q}}\right)=1$ and thereupon $\det\left(\mathbf{Q'}\right)=1$, which completes the proof.

\pagestyle{empty}

\end{document}